# New Variable Stars Discovered by the APACHE Survey. I. Results After the First Observing Season

**Mario Damasso**
*INAF-Astrophysical Observatory of Torino, Via Osservatorio 20, I-10025 Pino Torinese, Italy; Astronomical Observatory of the Autonomous Region of the Aosta Valley, fraz. Lignan 39, 11020 Nus (Aosta), Italy; Dept. of Physics and Astronomy, University of Padova, Vicolo dell'Osservatorio 3, I-35122 Padova, Italy; mario.damasso@studenti.unipd.it and m.damasso@gmail.com*

**Andrea Bernagozzi**
**Enzo Bertolini**
**Paolo Calcidese**
**Albino Carbognani**
**Davide Cenadelli**
*Astronomical Observatory of the Autonomous Region of the Aosta Valley, fraz. Lignan 39, 11020 Nus (Aosta), Italy*

**Jean Marc Christille**
*Dept. of Physics, University of Perugia, Via A. Pascoli, 06123 Perugia, Italy; Astronomical Observatory of the Autonomous Region of the Aosta Valley, fraz. Lignan 39, 11020 Nus (Aosta), Italy*

**Paolo Giacobbe**
*INAF-Astrophysical Observatory of Torino, Via Osservatorio 20, I-10025 Pino Torinese, Italy; Dept. of Physics, University of Trieste, Via Tiepolo 11, I-34143 Trieste, Italy*

**Luciano Lanteri**
**Mario G. Lattanzi**
**Richard Smart**
**Allesandro Sozzetti**
*INAF-Astrophysical Observatory of Torino, Via Osservatorio 20, I-10025 Pino Torinese, Italy*

**Abstract**   We present more than 80 new variable stars discovered during the first observing season of the APACHE survey. APACHE is a project aimed at detecting extrasolar planets transiting nearby, bright M dwarfs by using an array of small-aperture telescopes. Despite the fact that the survey is targeted to a well-defined sample of cool stars, we also reduce and analyze data for all the detected field stars. Since July 2012 dozens of different stellar fields have been monitored, leading to the detection of several variables for which we propose a classification and estimate a period, when a periodicity is evident in the data. Thanks to the SuperWASP public archive, we have also retrieved and analyzed photometric data collected by the SWASP survey, which helped us to refine the classification and the period estimation of many variables found in the APACHE database. Some of the variables present peculiarities and thus are discussed separately.

## 1. Introduction

Project APACHE (A PAthway toward the Characterization of Habitable Earths) is a new ground-based photometric survey specifically designed to search for transiting, small-size planets orbiting bright, nearby early-to-mid M dwarfs (Sozzetti *et al.* 2013). The project has been active, since July 2012, at the Astronomical Observatory of the Autonomous Region of the Aosta Valley (OAVdA), located in the Western Italian Alps.

APACHE (http://apacheproject.altervista.org) is a collaboration between OAVdA and INAF-Osservatorio Astrofisico di Torino (INAF-OATo), and it will last for five years. The survey utilizes an array of five automated 40-cm telescopes to monitor hundreds of M dwarfs properly selected from the catalogue of Lépine and Gaidos (2011). In parallel with the search for transit-like signals in the light curves of the M dwarfs, we are also analyzing the photometric data of the huge sample of stars that fall in the fields of view of the telescopes, which are centered on the target cool stars, to look for new variable stars. Here we report a list of variables discovered during the first observing season of APACHE. As demonstrated after the first year of operation, the large photometric database collected by APACHE represents a treasure trove for identifying new variable stars which will be constantly scoured.

## 2. Instruments and methods

The APACHE survey uses an array composed of identical Carbon Truss 40-cm f/8.4 Ritchey-Chrétien telescopes, each with a GM2000 10MICRON German mount and equipped with a FLI Proline PL1001E-2 CCD Camera and Johnson-Cousins V, R, and I filters. This instrument configuration is characterized by a pixel scale of 1.5 arcsec/pixel and a field of view of $26' \times 26'$. Except for a small number of targets which are monitored in the V band, all the fields are observed using the $I_c$ filter. The complete catalogue of M dwarfs eligible for observations by the APACHE survey is composed of 3,323 targets and has been organized in a ranking list by assigning observing priorities which take into account several factors, for example: the highest priority is given to stars with a reliable spectral classification published in the literature; to those that are known to be slow rotators, based on measurements of their projected rotational velocity v *sin* i; whether they have a low chromospheric activity as measured by the equivalent width of the H-α line and are not known as bright X-ray sources. A slow rotator and a star with a low activity level has to be preferred because, for the spectroscopic follow-up required to measure the mass of a transiting planetary candidate, the measurements of the radial velocity variations are less affected by the line broadening and the intrinsic jitter due to stellar activity.

It is interesting to mention another parameter considered for the definition of the ranking list: the expected number of observations of each target guaranteed by the Gaia satellite, which has been successfully launched and is expected to provide a very relevant contribution to extrasolar science. According to the scanning law of Gaia, it is possible to calculate the number of expected transits over a certain field after the nominal five years of mission. When building the APACHE schedule, we have assumed that targets with a number of Gaia observations greater than 100 have higher priorities because they will be characterized by a very accurate measurements of trigonometric parallax, from which it will be possible to accurately determine fundamental stellar parameters such as the intrinsic luminosity, mass, and radius, which are necessary for a precise study of any hosted planet.

The APACHE observations are carried out in focus, and the exposure times—which are kept fixed during the sessions independently from the seeing—are in the range of 3 to 180 seconds, while the magnitudes of the target M dwarfs vary within 8–16.5 in the V-band and 5.5–13 in the J-band. The APACHE telescopes use a circular observing schedule, with each target being re-pointed typically 20 minutes after the last observation. This cadence should be optimal for collecting enough data points which fall in the portion of the light curve showing a transit, if it is in progress, that usually is expected to last for one to three hours. Each time a target is pointed, three consecutive exposures are taken and usually the average value of the corresponding differential magnitudes is used for light curve analysis. In the course of a night the same field is observed as long as its altitude above the horizon is greater than 30 degrees, while it remains in schedule for typically 60 to 90 days in a row.

After the first observing season nearly 150 different stellar fields have been observed, necessarily characterized by not uniform amounts of data. We have searched for new variable stars especially considering those fields with the major amounts of photometric data. For each field we have visually inspected the differential light curves of all the stars identified, which typically amount to some dozens but in some cases are more than one thousand per field, and processed by the data reduction and analysis pipeline TEEPEE (Transiting ExoplanEt PipElinE), developed by the authors mostly in IDL programming language and tailored specifically to provide the user with optimal light curves of the primary target M dwarfs. The TEEPEE package, indifferently applied to data collected during a single night or over the whole timespan of observations, tests up to twelve different apertures and the best set of comparison stars is automatically determined from a list of dozens among the brightest objects in the field, excluding those too close to the CCD borders. The best aperture and set of reference stars selected at the end of the data processing are the ones which give the smallest RMS for the entire light curve of the object of interest. To save CPU time, this procedure is applied once with regard only to the primary M dwarf targets. Thus the aperture and set of comparisons used for determining the differential light curves of the field stars are those selected for the M dwarfs, which usually represent a good choice. Because, as stated, the number of comparison stars is typically high for each field (more than 10 in several cases), we do not provide the complete list for each variable we discuss here. At our site the seeing conditions can vary sensibly from night to night, and typical values for the FWHM of the APACHE images range within 1.5 to 3 pixels, and the TEEPEE pipeline tests aperture radii in the range 3.5 to 9 pixels.

## 3. Results

The aim of this work is to provide a list of new variable stars for which we propose tentative classification and—for those showing evidence of periodicity—we produce an estimate of the period. Our conclusions originate only from photometry, and no spectroscopic follow-up has been carried out to improve our knowledge about these objects. For some stars, we considered it interesting to provide a detailed, separate discussion (section 4). The complete list of 86 variables discussed in this work is presented in Table 1. All of them have been searched in the AAVSO VSX database and 83 can be considered as new genuine discoveries. Three objects are already included in the VSX database. One eclipsing binary system is provided without any information about the orbital period, which we have estimated from our data. Another object is classified in VSX as an eclipsing variable without any information about the orbital period, but our data do not confirm this classification, rather the star appears to be an irregular variable of type L. A variable of L type, as read in the VSX database, is reported with a period of 88 days, but this value is not confirmed by our data. In Figure 1 we show the corresponding light curves, with those showing periodicity folded according to the best period found through a Lomb-Scargle (L-S) analysis applied to the APACHE data, except in very few cases (as explained below). For this task we used the IDL version of the L-S algorithm (http://www.arm.ac.uk/ csj/idl/PRIMITIVE/scargle.pro). Almost every APACHE light curve presented here has been obtained using the $I_c$ filter, except where specified in the explanatory notes accompayining Table 1. The magnitudes in V band, where not otherwise specified, are those from the APASS survey, which are reported by the UCAC4 catalogue (Zacharias *et al.* 2012). Several stars appear as having no evidence of periodicity, or with a very poorly defined periodicity which shows only occasionally. Due to this

Table 1. Variable stars discovered by APACHE during the first observing season. For stars with two periods indicated, that in parentheses is the one determined from the APACHE data, while the other is estimated from SuperWASP archive data. The amplitudes of the light curves are measured from APACHE data. Time $T_0$ corresponds to phase = 0 in the folded light curves.

| No. | Name | R.A. (2000) | Dec. (2000) | Mag V | Period (days) | Amplitude (mag) | $T_0$ (HJD–2455000) | Var. Type | Note |
|---|---|---|---|---|---|---|---|---|---|
| 1 | UCAC4 744-001518 | 002.6931077 | +58.6863575 | 14.86 | 7.55 (7.518) | 0.25 | 1146.3523197 | ROT? | (1) |
| 2 | UCAC4 743-001636 | 002.7103159 | +58.5618373 | 10.21 | 0.10625 (0.09603) | 0.02 | 1146.3523197 | DSCT | (2) |
| 3 | UCAC4 758-009639 | 015.1541895 | +61.5753528 | 14.27 | 1.20557 | 0.6 | 1146.4105864 | E | (3) |
| 4 | UCAC4 752-014548 | 018.8161518 | +60.3735250 | 13.14 | 1.087 | 0.3 | 1146.3866004 | EB | |
| 5 | UCAC4 870-000885 | 018.9877518 | +83.9571120 | 16.77 | 0.40861 | 0.3 | 1273.2157609 | EW | |
| — | UCAC4 872-000839 | 021.2561924 | +84.2626639 | 16.17 | ? | — | | RR? | (4) |
| 6 | UCAC4 728-026837 | 044.2308518 | +55.5203950 | 16.38 (GSC2.3) | ? | 0.4 | 1146.4709166 | L | |
| 7 | UCAC4 618-013561 | 062.4040074 | +33.4937250 | 12.83 | ? | 0.25 | 1162.5960791 | L | |
| 8 | UCAC4 713-031969 | 063.5650233 | +52.5300092 | 14.4 | 6.22 | 0.4 | 1208.3673785 | E | (5) |
| 9 | UCAC4 709-034533 | 074.7646086 | +51.7157109 | 10.77 | ? | 0.05 | 1328.2561929 | L | |
| 10 | UCAC4 610-021265 | 082.4603353 | +31.9110003 | 13.59 | 0.80089 (0.80086) | 0.3 | 1210.5113093 | EW | (6) |
| 11 | UCAC4 619-030850 | 092.0415889 | +33.6711831 | 13.01 | 2.008 (2.009) | 0.27 | 1205.5239017 | EB | (7) |
| 12 | UCAC4 521-023140 | 094.9127039 | +14.0543662 | 12.47 | ~77? | 0.16 | 1270.3904921 | L | (8) |
| 13 | UCAC4 624-036803 | 105.7816774 | +34.7009262 | 15.90 | 0.21635 (WASP) | 0.25 | 1246.5895718 | EW | (9) |
| 14 | UCAC4 715-044844 | 105.9254200 | +52.8823920 | 13.22 | 0.3281 (0.32811) | 0.33 | 1205.4926955 | EW | (10) |
| 15 | UCAC4 664-056989 | 179.4896992 | +42.6626264 | 14.19 | 0.33651 | 0.5 | 1353.4968403 | RRab | (11) |
| 16 | UCAC4 609-049172 | 208.1377689 | +31.6871487 | 13.47 | 0.35471 (0.35470) | 0.16 | 1301.7535576 | EW | (12) |
| 17 | UCAC4 666-060394 | 232.1916465 | +43.0916548 | 12.79 | 1.9756 (1.9788) | 0.1 | 1127.3672611 | ROT | (13) |
| 18 | UCAC4 633-053996 | 250.3480433 | +36.5047900 | 12.26 | ? | 0.1 | 1367.4885131 | L | |
| 19 | UCAC4 633-054311 | 250.4026556 | +36.4431934 | 15.06 | ? | 0.3 | 1367.4885131 | L | (14) |

Table 1. Variable stars discovered by APACHE during the first observing season. For stars with two periods indicated, that in parentheses is the one determined from the APACHE data, while the other is estimated from SuperWASP archive data. The amplitudes of the light curves are measured from APACHE data. Time $T_0$ corresponds to phase =0 in the folded light curves, cont.

| No. | Name | R.A. (2000) | Dec. (2000) | Mag V | Period (days) | Amplitude (mag) | $T_0$ (HJD-2455000) | Var. Type | Note |
|---|---|---|---|---|---|---|---|---|---|
| 20 | UCAC4 633-054908 | 250.4621462 | +36.4817620 | 11.82 | ? | 0.2 | 1367.4885131 | L | (15) |
| 21 | UCAC4 652-057450 | 253.3608262 | +40.3033556 | 14.83 | 1.47 | 0.45 | 1420.4785809 | EB | |
| 22 | UCAC4 592-063874 | 275.7670439 | +28.3080614 | 14.01 | ~60? | 60.2 | 1127.3682196 | L | |
| 23 | UCAC4 533-077928 | 279.6991892 | +16.4103631 | 15.12 | ? | 60.5 | 1166.3294430 | L | |
| 24 | UCAC4 533-078069 | 279.8361324 | +16.5350567 | 14.97 | ? | 60.2 | 1166.3294430 | L | |
| 25 | UCAC4 532-076755 | 279.8824256 | +16.3173617 | 15.15 | 0.3199 | 60.25 | 1166.3294430 | EW | |
| 26 | UCAC4 533-078283 | 280.0367009 | +16.5274503 | 14.58 | ? | 60.11 | 1166.3294430 | L | |
| 27 | UCAC4 723-061541 | 283.8471103 | +54.4308875 | 14.33 | 0.38399 (0.38399) | 60.7 | 1385.4985908 | EW | (16) |
| 28 | UCAC4 723-061622 | 284.1744556 | +54.5157323 | 13.34(GSC2.3) | ? 6 | 0.35 | 1385.4876350 | L | |
| 29 | UCAC4 479-089263 | 284.6627689 | +05.7352287 | 11.68 | 0.744 | 60.1 | 1473.4154161 | RRc | (17) |
| 30 | UCAC4 491-099556 | 284.8352648 | +08.1355764 | N.A. | ? | 60.15 | 1432.4879724 | L | (18) |
| 31 | 2MASS 18592325+0810247 | 284.846896 | +08.1735555 | N.A. | ? | 60.25 | 1432.4879724 | L | (19) |
| 32 | UCAC4 491-099593 | 284.8855603 | +08.0666542 | 16.48(GSC2.3) | 1.251 | 60.35 | 1432.4879724 | EW | |
| 33 | 2MASS 19000738+0805125 | 285.030765 | +08.086826 | 19.00(GSC2.3) | ? | 0.25 | 1432.4879724 | L | |
| 34 | UCAC4 490-094834 | 285.0720153 | +07.9374948 | 16.58(GSC2.3) | ? | 60.12 | 1432.4879724 | L | |
| 35 | UCAC4 729-060138 | 290.8229630 | +55.6600275 | 12.76 | 0.6902 (0.688) | 60.08 | 1434.3738137 | ROT | (20) |
| 36 | UCAC4 728-061629 | 290.9937003 | +55.5546067 | 14.81 | 0.36689 (0.3669) | 60.2 | 1434.3728183 | EW | (21) |
| 37 | UCAC4 728-061766 | 291.4820559 | +55.5007378 | 15.27(GSC2.3) | 0.189029 (0.1890) | 60.1 | 1434.3738137 | DSCT | (22) |

*Table continued on following pages*

Table 1. Variable stars discovered by APACHE during the first observing season. For stars with two periods indicated, that in parentheses is the one determined from the APACHE data, while the other is estimated from SuperWASP archive data. The amplitudes of the light curves are measured from APACHE data. Time $T_0$ corresponds to phase =0 in the folded light curves, cont.

| No. | Name | R.A. (2000) | Dec. (2000) | Mag V | Period (days) | Amplitude (mag) | $T_0$ (HJD–2455000) | Var. Type | Note |
|---|---|---|---|---|---|---|---|---|---|
| 38 | UCAC4 811-027074 | 294.4105074 | +72.0363334 | 13.63 | ? | 70.2 | 1424.3380554 | L | |
| 39 | UCAC4 610-092144 | 297.3880989 | +31.8705942 | 14.72(NOMAD) | ? | 70.1 | 1127.5857519 | L | |
| 40 | UCAC4 609-091131 | 297.4135403 | +31.6445878 | 14.79(NOMAD) | ? | 70.15 | 1127.5857519 | L | |
| 41 | UCAC4 609-091297 | 297.4814474 | +31.7204662 | 17.27(NOMAD) | ? | 70.9 | 1127.5857519 | L | |
| 42 | 2MASS | | | | | | | | |
| | J19500382+3149132 | 297.515956 | +31.820337 | N.A. | ? | 70.15 | 1127.5857519 | L | |
| 43 | 2MASS | | | | | | | | |
| | J19500885+3135499 | 297.536909 | +31.597216 | N.A. | ? | 70.4 | 1127.5857519 | L | |
| 44 | UCAC4 610-092592 | 297.5731206 | +31.9941723 | 15.53(NOMAD) | ? | 70.2 | 1127.5857519 | L | |
| 45 | 2MASS | | | | | | | | |
| | J19502059+3152091 | 297.585806 | +31.869217 | 17.38(NOMAD) | ? | 70.2 | 1127.5857519 | L | |
| 46 | 2MASS | | | | | | | | |
| | J19502935+3158417 | 297.622323 | +31.978273 | N.A. | ? | 70.17 | 1127.5857519 | L | |
| 47 | UCAC4 610-092815 | 297.6690953 | +31.9639217 | 15.41(NOMAD) | 0.427 | 70.5 | 1127.5857519 | EW | |
| 48 | UCAC4 609-091797 | 297.7097806 | +31.6965225 | 12.70 | ~0? | 70.13 | 1127.5857519 | L | |
| 49 | UCAC4 608-095594 | 297.7998224 | +31.5728037 | 15.22(NOMAD) | ? | 70.45 | 1127.5857519 | L | |
| 50 | 2MASS | | | | | | | | |
| | J19511471+3143128 | 297.811294 | +31.720238 | 17.47 | ? | 70.14 | 1127.5857519 | L | |
| 51 | UCAC4 623-096266 | 302.9632515 | +34.4131084 | 15.28(NOMAD) | 9.8 | 70.45 | 1416.5074270 | PULS | (23) |
| 52 | UCAC4 623-096337 | 303.0224745 | +34.4758434 | 15.87(NOMAD) | 761? | 70.16 | 1416.5074270 | L | |

*Table continued on following pages*

Table 1. Variable stars discovered by APACHE during the first observing season. For stars with two periods indicated, that in parentheses is the one determined from the APACHE data, while the other is estimated from SuperWASP archive data. The amplitudes of the light curves are measured from APACHE data. Time $T_0$ corresponds to phase =0 in the folded light curves, cont.

| No. | Name | R.A. (2000) | Dec. (2000) | Mag V | Period (days) | Amplitude (mag) | $T_0$ (HJD-2455000) | Var. Type | Note |
|---|---|---|---|---|---|---|---|---|---|
| 53 | UCAC4 621-098928 | 303.0456662 | +34.0879998 | 13.90 | 0.1365 | 80.07 | 1416.5076083 | DSCT? | |
| 54 | UCAC4 622-095023 | 303.0508565 | +34.2564675 | 14.06(NOMAD) | ? | 80.22 | 1416.5074270 | L | (24) |
| 55 | UCAC4 622-095081 | 303.0868139 | +34.3034262 | 17.07(NOMAD) | 844? | 80.15 | 1416.5074270 | L | |
| 56 | UCAC4 622-095314 | 303.2126174 | +34.2968739 | 16.19(NOMAD) | ? | 80.12 | 1416.5074270 | L | |
| 57 | UCAC4 623-096673 | 303.2543371 | +34.4226409 | 14.72(NOMAD) | ? | 80.25 | 1416.5074270 | L | |
| 58 | UCAC4 622-095521 | 303.3453327 | +34.3789689 | 13.58 | ? | 80.3 | 1416.5074270 | L | |
| 59 | UCAC4 622-095554 | 303.3731539 | +34.3134381 | 14.99(NOMAD) | ? | 80.05 | 1416.5074270 | L | |
| 60 | UCAC4 516-127264 | 303.6096042 | +13.1699259 | 12.40 | 0.575 | 80.2 | 1445.5441597 | EW | |
| — | UCAC4 744-062741 | 306.3905642 | +58.6538473 | 14.09 | 1.51? | 80.2 | 1127.3642954 | ROT? | (25) |
| 61 | UCAC4 744-062753 | 306.4085277 | +58.7603659 | 14.17(GSC2.3) | 0.4406 (0.4405) | 80.25 | 1127.3642954 | EW | (26) |
| 62 | UCAC4 744-062788 | 306.5188756 | +58.7358845 | 15.2(GSC2.3) | 0.3495 | 80.45 | 1127.3642954 | EW | |
| 63 | 2MASS 2030S052+5547074 | 307.710540 | +55.785416 | 18.64(GSC2.3) | ? | 82.0 | 1417.5662968 | L | |
| 64 | UCAC4 729-067019 | 307.8310009 | +55.7983139 | 15.56(GSC2.3) | ? | 80.6 | 1417.5664087 | L | |
| 65 | UCAC4 731-068791 | 308.1278424 | +56.1299364 | 14.44 | ? | 80.4 | 1417.5664087 | L | |
| 66 | UCAC4 621-112859 | 314.4579859 | +34.1641767 | 13.84(NOMAD) | ? | 80.15 | 1432.5597351 | L | (27) |
| 67 | UCAC4 617-116284 | 315.4994471 | +33.3210739 | 14.90(NOMAD) | ? | 80.5 | 1127.3687558 | L | |
| 68 | UCAC4 598-126361 | 317.6277274 | +29.4905828 | 12.48 | ? | 80.2 | 1443.4864670 | L | |
| 69 | UCAC4 621-122572 | 321.8936342 | +34.0494003 | 12.09 | 0.43635 (0.438) | 80.1 | 1445.5736404 | EW | (28) |
| 70 | UCAC4 590-130214 | 326.9765700 | +27.9013675 | 12.71 | 836 | 80.2 | 1135.4185368 | L | (29) |

*Table continued on following pages*

Table 1. Variable stars discovered by APACHE during the first observing season. For stars with two periods indicated, that in parentheses is the one determined from the APACHE data, while the other is estimated from SuperWASP archive data. The amplitudes of the light curves are measured from APACHE data. Time $T_0$ corresponds to phase = 0 in the folded light curves, cont.

| No. | Name | R.A. (2000) | Dec. (2000) | Mag V | Period (days) | Amplitude (mag) | $T_0$ (HJD–2455000) | Var. Type | Note |
|---|---|---|---|---|---|---|---|---|---|
| 71 | UCAC4 590-130270 | 327.0770353 | +27.8228681 | 12.49 | 1.02500 (1.025) | 90.25 | 1127.3844791 | EW | (30) |
| 72 | UCAC4 652-105561 | 327.4152459 | +40.3802687 | 14.33 | ? | 90.35 | 1443.4839005 | L | |
| 73 | UCAC4 588-128603 | 330.1270789 | +27.4577517 | 12.72 | 4.21221 (1.29) | 90.08 | 1459.5047071 | ROT? | (31) |
| 74 | UCAC4 789-036290 | 330.5805324 | +67.6531512 | 16.00(GSC2.3) | 0.17 (0.34) | 90.15 | 1127.3607291 | DSCT(EW) | (32) |
| 75 | UCAC4 788-037466 | 330.9508409 | +67.4961639 | 14.29(GSC2.3) | 0.47 | 90.4 | 1127.3602622 | EA | |
| 76 | UCAC4 788-037537 | 331.1173995 | +67.4969889 | 13.66(GSC2.3) | ? | 90.16 | 1127.3602622 | L | |
| 77 | UCAC4 788-037577 | 331.2463553 | +67.5745589 | 15.62(GSC2.3) | 0.36 | 90.25 | 1127.3607291 | EW | |
| 78 | UCAC4 726-083454 | 333.0765986 | +55.1851717 | 14.67 | ? | 90.18 | 1127.3607291 | L | |
| 79 | UCAC4 725-086173 | 333.2373680 | +54.9382473 | 13.34 | 0.192835 | 90.2 | 1127.3607291 | DSCT? | |
| 80 | UCAC4 725-086241 | 333.2829348 | +54.8638348 | 14.33 | 0.17234 | 90.5 | 1443.4626241 | DSCT? | |
| 81 | UCAC4 781-040386 | 333.4526621 | +66.1133137 | 15.43 | 1.13 | 90.2 | 1127.3604366 | EB | |
| 82 | UCAC4 726-084171 | 333.5242624 | +55.0744792 | 12.13 | 950 | 90.15 | 1127.3604366 | L | |
| 83 | UCAC4 859-013147 | 343.6693409 | +81.7063873 | 12.72 | 37 | 90.17 | 1385.653719 | L | |
| 84 | UCAC4 749-082493 | 352.8942900 | +59.7192198 | 10.64 | 0.861 | 90.2 | 1146.4200537 | EW | |

Table 1. Variable stars discovered by APACHE during the first observing season, cont.

*Notes: (1) Star 1SWASP J001046.35+584111.0, 1837 points analyzed. APACHE data are less scattered than those of SWASP. (2) Star 1SWASP J001050.47+58334 2.6, SWASP data with high scatter, 1969 of 2015 data analyzed (applying a 1-sigma clipping to the original data). (3) Eclipsing binary already known in VSX (Mis V1368) but the orbital period was not reported. (4) The light curve is discussed separately in Figure 2. No SWASP data available for this star. (5) Only the primary minimum detected. (6) Star 1SWASP J052950.48+315439.8, 7033 of 7110 data points from SWASP analyzed (applying a 3-sigma clipping to the original data). The secondary minimum appears deeper in the light curve from the APACHE survey. Refer to Figure 3 for a comparison between SWASP and APACHE light curves. (7) Star 1SWASP J060809.97+334016.5, 6223 of 6282 data points from SWASP analyzed (applying a 3-sigma clipping to the original data). The existence of a secondary minimum can be guessed from the APACHE light curve, while is not visible in SWASP data (not shown here). (8) Tentative period, but the phase coverage does not make possible a reliable estimation. (9) Star 1SWASP J070307.60+344203.60, 4066 of 4305 data points analyzed (applying a 2-sigma clipping to the original data). Even with few data points, this faint star appears to be an eclipsing binary system in the APACHE photometry. From data in the SWASP public archive we derive the best period P=0.21635 days, that in this case we adopted for folding our data because is a far better estimate than that from APACHE data. This star is discussed separately in section 4. (10) Star 1SWASP J070342.10+525256.8, noisy light curve, 1094 of 1258 data points from SWASP analyzed (applying a 1-sigma clipping to the original data). (11) Star 1SWASP J115757.51+423945.5, 5509 data points analyzed. From APACHE photometry alone, because the data are few, we could only guess that the star has a short period (<1 day), but a reliable determination was not possible. SWASP photometry helped us in classifying the star's variability and estimating a reliable period. We show part of the APACHE light curve and the SWASP data in Figure 5. (12) Star 1SWASP J135233.09+314113.8, 7348 of 7425 data points analyzed (applying a 3-sigma clipping to the original data). (13) Star 1SWASP J152845.99+430530.1, 5387 of 5428 data points analyzed (applying a 3-sigma clipping to the original data). Spotted star. Color indexes B–V = 0.98, V–J = 1.78, V–K = 2.39. Tentative spectral classification: dK3/dK4. (14) Appears in Kopacki et al. (2003). V from Sandquist et al. (2010). (15) Appears in Kopacki et al. (2003). (16) Star 1SWASP J185523.32+542551.4, 11972 of 12016 data points analyzed (applying a 3-sigma clipping to the original data). (17) Observations in V band. Color indexes: B–V = 0.95, V–J = 1.7, V–K = 2.33. (18) In on-line archive images it appears as a blended object. Faint, no V available. (19) Faint object in on-line archive images. (20) Star 1SWASP J192317.5+553936.2, 12861 of 12919 data points analyzed (applying a 3-sigma clipping to the original data). Results for this star are discussed separately in section 4. (21) Star 1SWASP J192358.37+553316.9, 12469 of 12596 data points analyzed (applying a 3-sigma clipping to the original data). (22) Star 1SWASP J192555.70+553002.9, 12934 photometric points used in the analysis. (23) Undefined type of pulsating star. Modulation possibly due to rotation. We provide a more conservative uncertainty than that calculated as 1/timespan. (24) Star listed in the International Variable Star Index (NSVS J2012124+341522) and classified as L. The reported periodicity of 88 days is not found in our data that clearly show that the possible period should be longer. (25) Star 1SWASP J202533.73+583913.9, 10637 of 10721 data points analyzed (applying a 3-sigma clipping to the original data). Because of its uncertain classification, this star is discussed separately in section 4. (26) Star 1SWASP J202558.05+584537.4, 10485 of 10550 data points analyzed (applying a 3-sigma clipping to the original data). The primary minimum is not well sampled in the APACHE light curve. (27) Star listed in the International Variable Star Index (NSV 25408) and classified as an eclipsing binary, but without an estimate of the orbital period. No evidence for eclipses is present in the APACHE data. Rather, the star appears as an irregular variable. Thus we propose a change of variability status for this star. (28) Star 1SWASP J212734.47+340257.9, 14781 of 14872 data points analyzed (applying a 3-sigma clipping to the original data). (29) Long periodicity, assumed tentatively nearly equal to 36 days by looking at the light curve. (30) Star 1SWASP J214818.48+274922.0, 14086 of 14151 data points analyzed (applying a 3-sigma clipping to the original data). (31) Star 1SWASP J220030.52+272728.8, 21352 of 21499 data points analyzed (applying a 3-sigma clipping to the original data). This star is discussed in detail in section 4. Even if the classification remains uncertain, based on SWASP data we suggest this could be a rotating star. (32) Difficult classification, faint star/noisy light curve. We propose two possibilities, with corresponding periodicities.*

circumstance, we propose these stars to be classified as a variable of L type. The remainder of our list is composed of periodic variables, mostly eclipsing binary systems and some pulsators, with a classification not always clear. To improve the period  estimation and to provide a less ambiguous classification where required, we used the SuperWASP public archive—a precious treasure trove now hosted by the NASA Exoplanet Archive (http://exoplanetarchive. ipac.caltech.edu/index.html)—to recover the light curves of our stars. Nearly 25% of the stars in our sample have thousands of useful SWASP photometric data points that can complement the APACHE dataset with high profit. We performed a L-S analysis to the data from the SWASP survey, specifically using the magnitudes de-trended with the SysRem algorithm (Tamuz *et al.* 2005). In Table 1 we report out of parentheses the best period found from the SWASP data (when available) which usually, because the time span and number of data are greater, can be indicated with a number of significant digits higher than for the period derived from APACHE timeseries (reported in parentheses). In a few cases, we report only the period found from SWASP data, which has been used to obtain the plot in Figure 1. The position of the last significant decimal digit, both for SWASP and APACHE data, is determined in the following way. We start by folding the light curve using the best period found by the L-S algorithm. Then, after removing the last decimal digit, we fold the data with this truncated period and, by visual inspection, we look at any change in the new phased light curve. If no change is evident, this means that the removed digit is not significant, and we reiterate the process by removing the second to last digit, and so on. When a change in the folded light curve is seen, this means that the removed digit is significant and the process comes to an end.

## 4. Discussion of individual variables

A small subset of variables deserves closer investigation.

### 4.1. UCAC4 872-000839

The star UCAC4 872-000839 (the first non-numbered object in Table 1) showed in the APACHE photometry and on nightly basis a variability similar to that of a pulsating star that we could not characterize with a well defined period. Figure 2 shows a sample of APACHE light curves of this star collected during single nights. Unfortunately this star is not present in the SWASP archive.

### 4.2. UCAC4 624-036803

The quite faint star UCAC4 624-036803 (Number 13 in Table 1) appears as an eclipsing binary system already in the APACHE data, and we obtained the best estimate for the orbital period $P = 0.21635$ day from SWASP data. This value is just below the lower limit of ~ 0.22 day for such systems discussed in Norton *et al.* (2011). This circumstance makes that star an interesting object for follow-up studies, as we have done for a similar variable (Damasso *et al.* 2012). The highest peak in the L-S periodogram of SWASP data is P=0.108176 day, which we have doubled to get the most reliable periodicity. The folded SWASP light curve and the corresponding L-S periodogram in the region of interest are shown in Figure 4, while data from the APACHE survey are shown in Figure 1, folded according to the best period found from SWASP photometry.

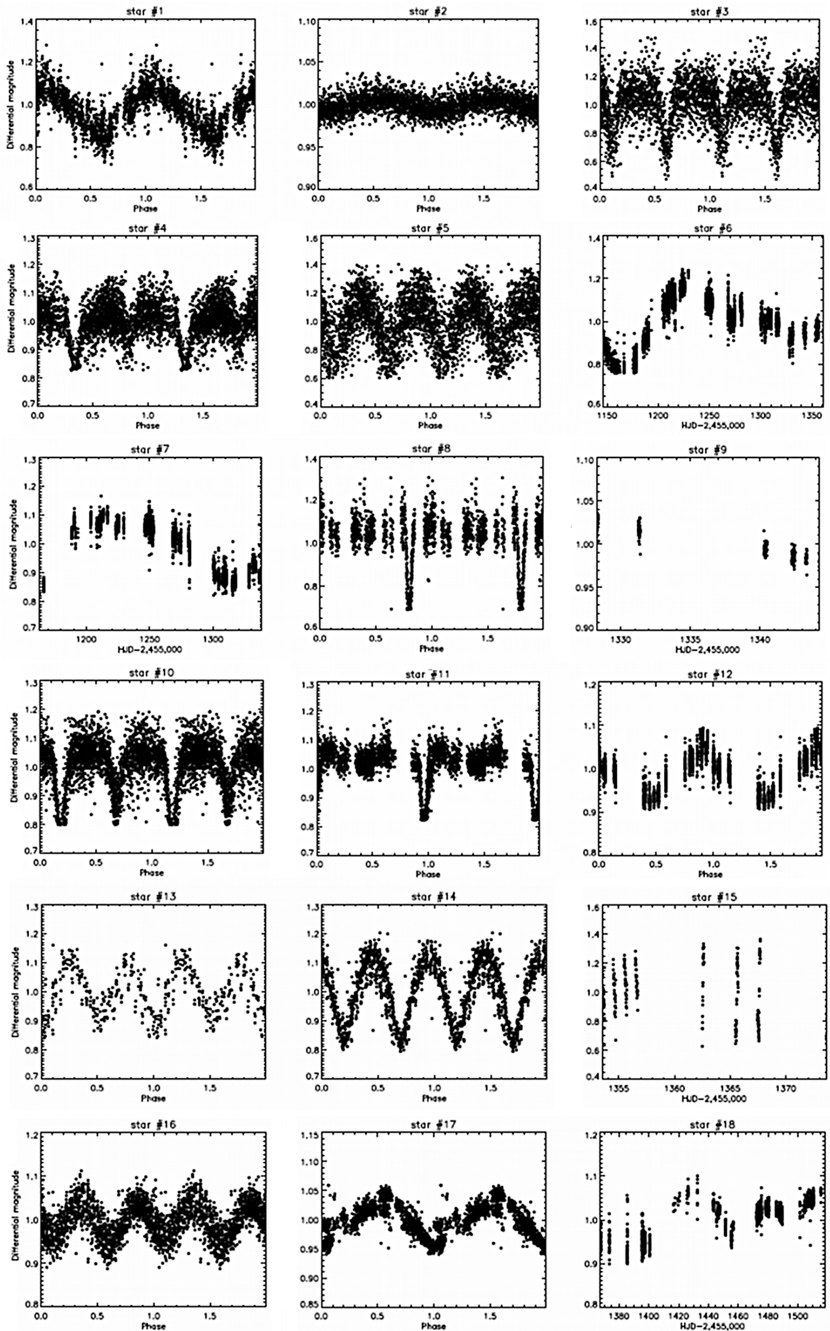

Figure 1. Light curves of the new variables listed in Table 1. They were found during the first observing season of the APACHE survey (figure continued on following pages).

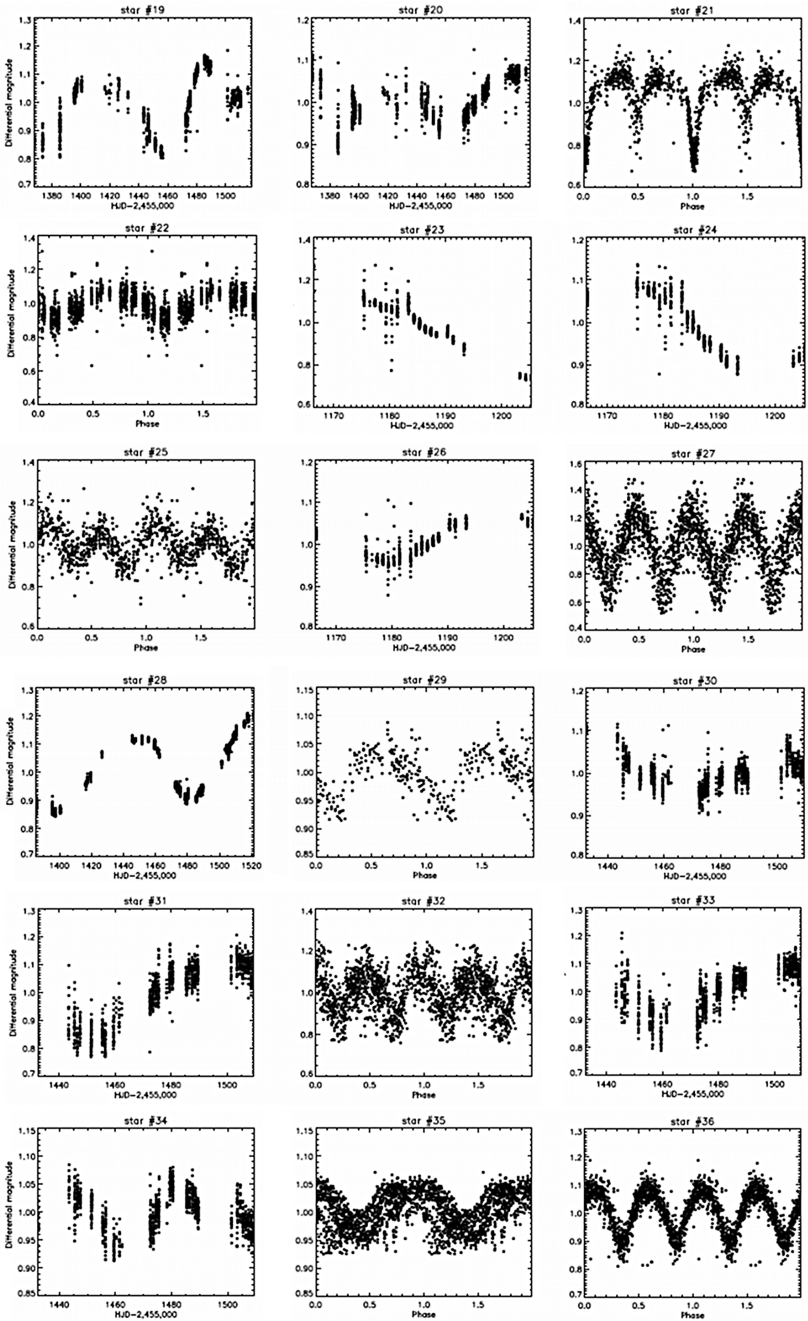

Figure 1. Light curves of the new variables listed in Table 1, continued.

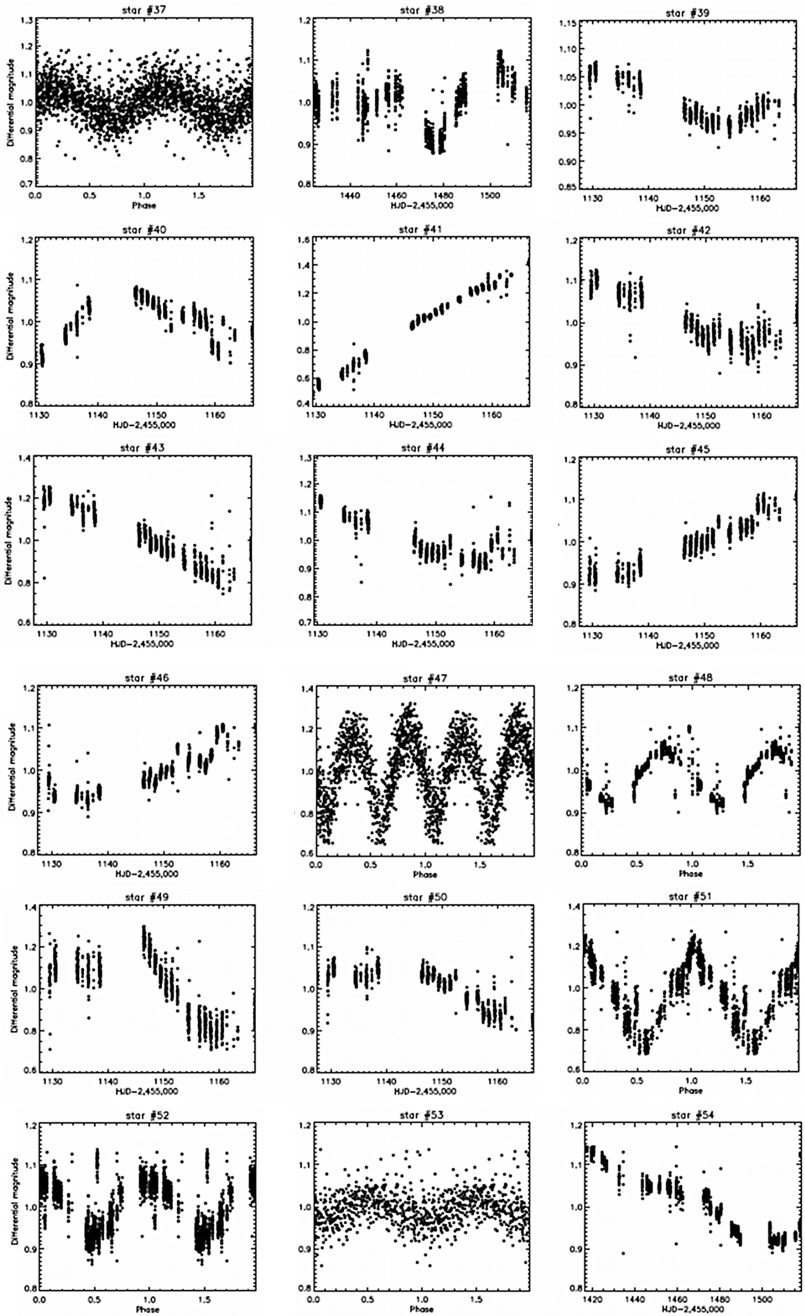

Figure 1. Light curves of the new variables listed in Table 1, continued.

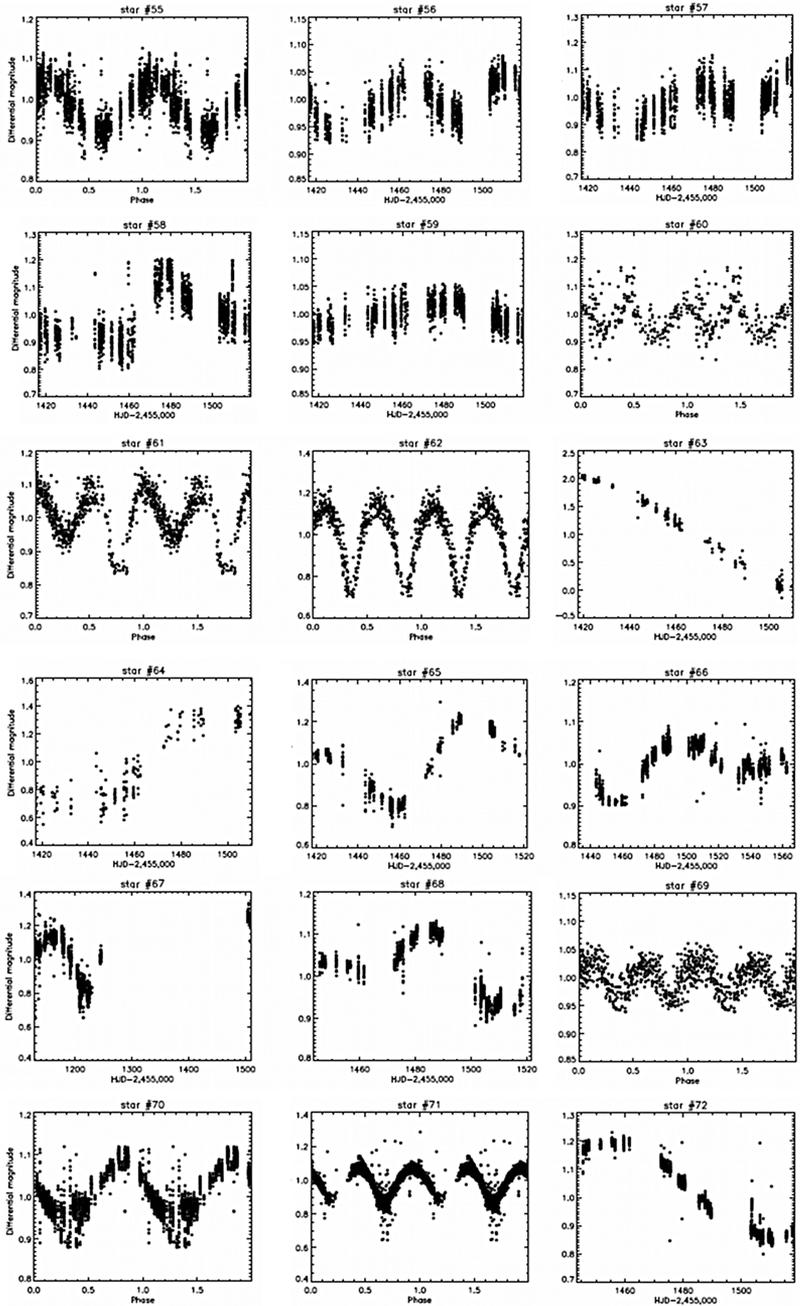

Figure 1. Light curves of the new variables listed in Table 1, continued.

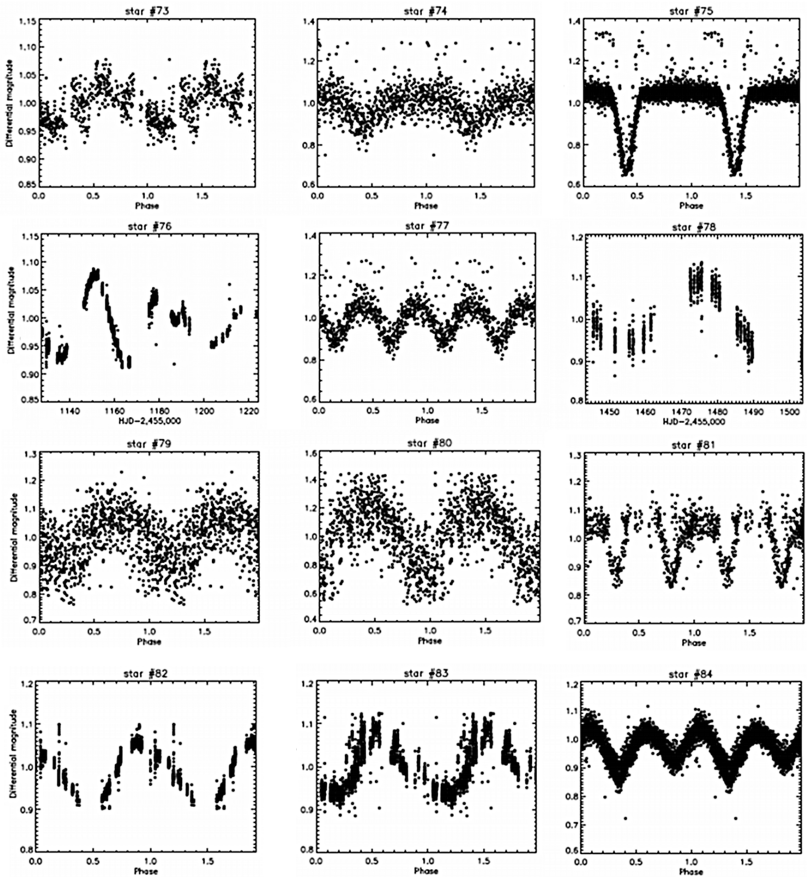

Figure 1. Light curves of the new variables listed in Table 1, continued.

### 4.3. UCAC4 664-056989

In the first plot of Figure 5, data for six nights are shown for the star UCAC4 664-056989 (Number 15 in Table 1). It has been observed by APACHE only for a few epochs, and it was not possible to determine a reliable periodicity. By looking at the single-night light curves it can be guessed that the star should be an RR Lyrae-type variable, but a clear classification is not possible. Luckily, this star has been observed by SWASP (1SWASPJ115757.51+423945.5), and we can quite safely propose classification as an RRab variable by looking at the SWASP light curve characteristics.

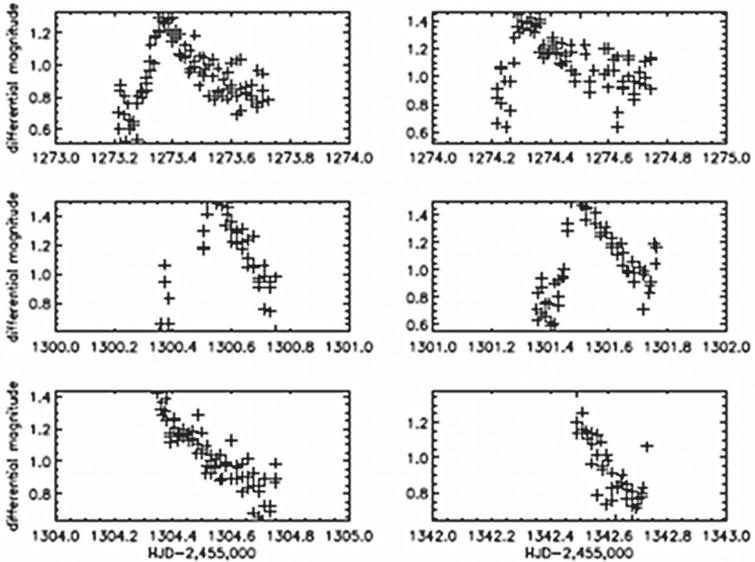

Figure 2. Sample of APACHE light curves of the star UCAC4 872-000839 obtained during six observing nights. Looking at the characteristics of the flux variations, the star can be tentatively and generically classified as a pulsating variable, but lacking a reliable determination of the periodicity. No data from the SWASP archive are available.

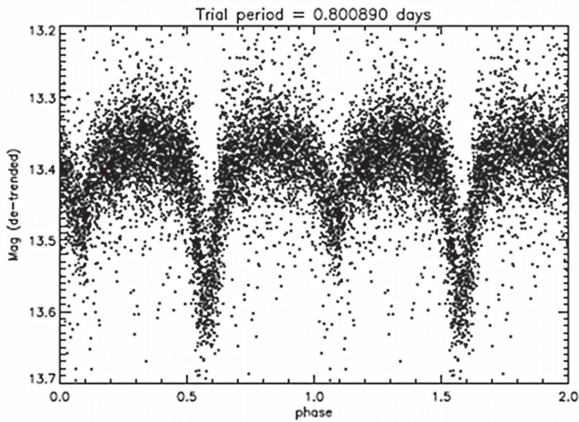

Figure 3. Light curve of the star UCAC4 610-021265 (Number 10 in Table 1) obtained with data from the WASP survey and folded according to the best period found P = 0.80089 day. By comparing this plot with that in Figure 1, it can be seen that the secondary minimum appears deeper in the APACHE data, obtained using an I$_c$ filter. See Note 6 in Table 1.

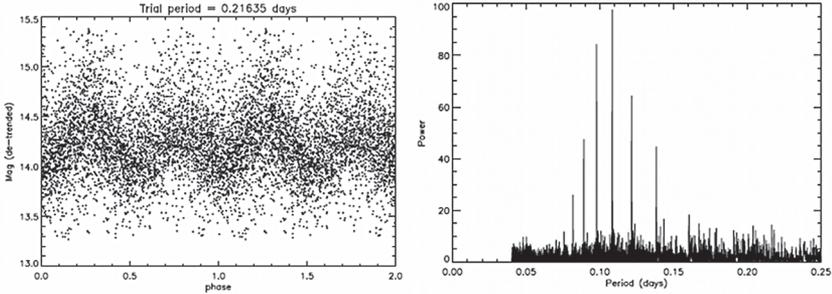

Figure 4. Left: light curve of the star UCAC4 624-036803 (SWASP J070307.60+344203.60; Number 13 in Table 1) obtained from data of the SuperWASP survey. Data are folded according to double the period with the highest peak in the Lomb-Scargle periodogram and rounded to the last significant digit. Right: the Lomb-Scargle periodgram in the range of periods of interest.

### 4.4. UCAC4 729-060138

The star UCAC4 729-060138 (Number 35 in Table 1) has B–V ≃1.0 according to the APASS photometry, and the dust reddening in the star's direction is estimated to be $E(B–V)=0.1$ (Schlafly and Finkbeiner 2011), thus suggesting that, assuming the star is a dwarf, it should be a late G to early K type. We propose that this variable should be classified as a rotator. Folding the APACHE data with the best period found (0.688 day) results in a light curve quite scattered in regard to the maximum, maybe indicating a change in amplitude that could be attributed to evolving active regions on the stellar surface. This star is also present in the SWASP database and a L-S analysis resulted in the best period of 0.6902 day. The corresponding folded light curve is showed in Figure 6. The curve from SWASP does not show the kind of amplitude scatter present in the APACHE data.

### 4.5. UCAC4 744-062741

The star UCAC4 744-062741 (second non-numbered object in Table 1) shows a clear (and with a complicated pattern) variability which was not possible to classify from APACHE photometry alone (first plot in Figure 7). The number of our data is not sufficient to clearly understand whether this is an eclipsing binary system or, generally speaking, a pulsating variable. The analysis of the SWASP light curve results in the most probable period P=1.5174 days, according to both the L-S and BLS algorithms (Kovács *et al.* 2002), and we show the folded data in the second panel of Figure 7. The analysis of the APACHE data, less numerous than those of SWASP, with the L-S algorithm reveals that the greatest power in the periodogram occurs at P=1.51 days (third

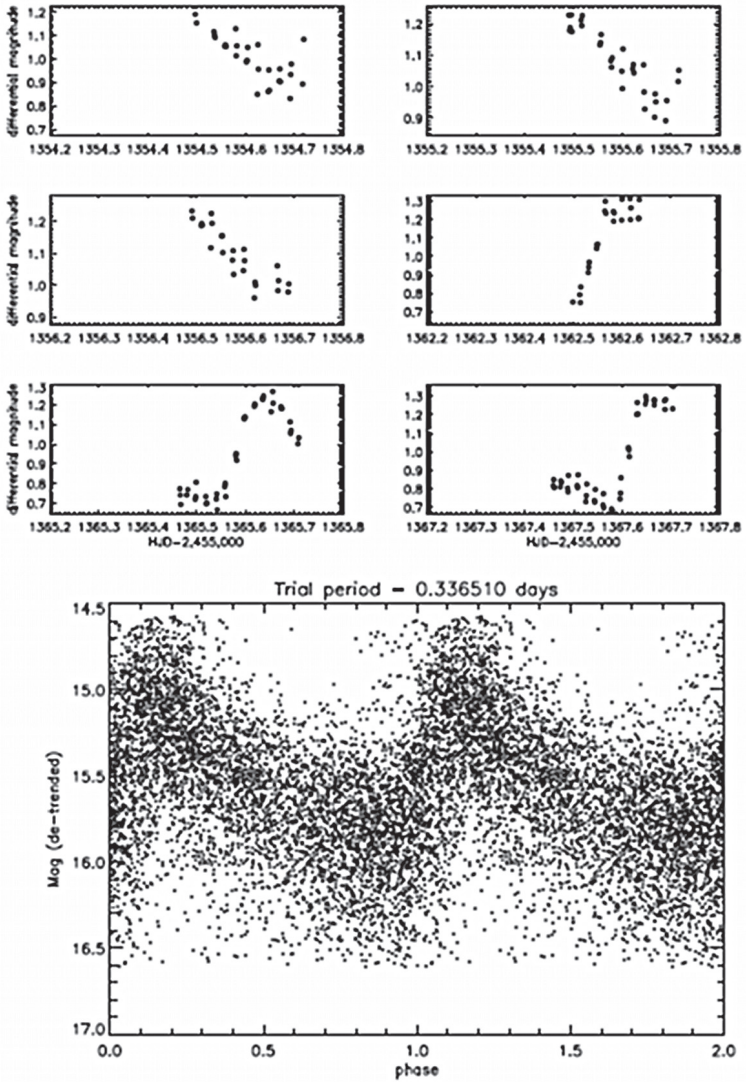

Figure 5. Top: sample of APACHE light curves of the star UCAC4 664-056989 obtained during six observing nights. Looking at the characteristics of the flux variations, the star can be tentatively and generically classified as an RRab variable, but lacking a reliable determination of the periodicity. Bottom: the real nature of this variable is unveiled, and a reliable period estimation made possible, thanks to the data in the SuperWASP archive, shown here. The star indeed appears to be an RRab variable.

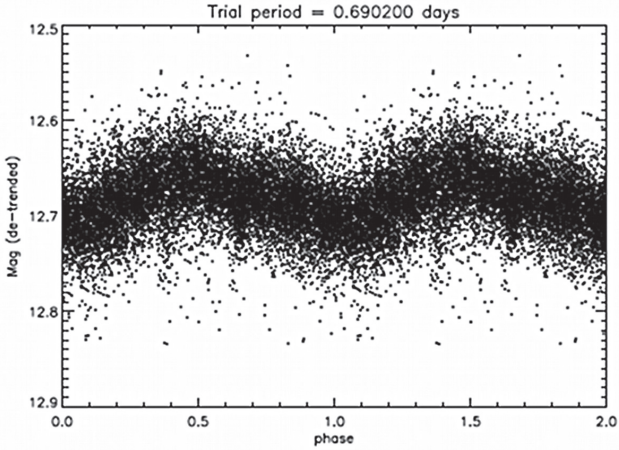

Figure 6. Light curve of the star UCAC4 729-060138 obtained with data from the WASP survey and folded according to the best period found P = 0.6902 day.

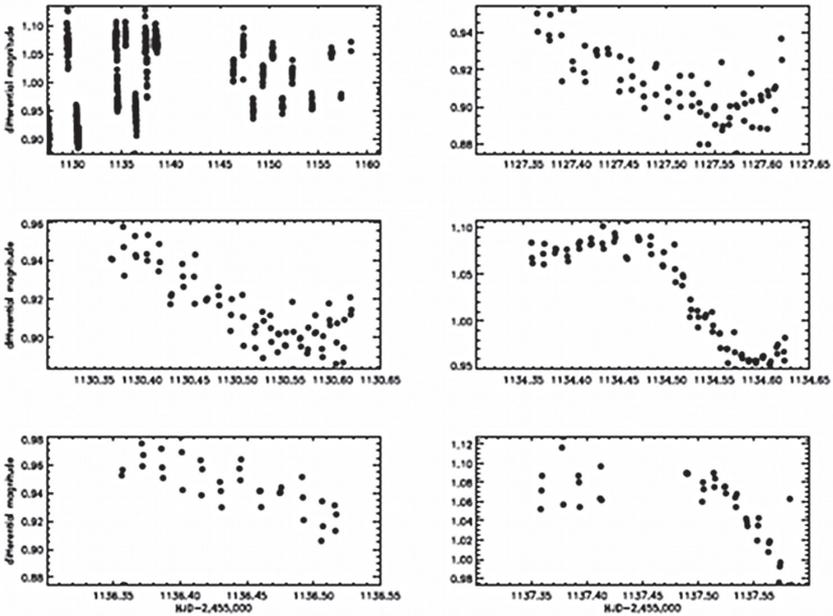

Figure 7. First plot (six panels): sample of APACHE light curves of the star UCAC4 744-062741. The upper left plot shows the light curve during the whole timespan of observations. The other plots show the light curves obtained during five different nights. They correspond to the 1st, 3rd, 4th, 6th, and 7th sub-intervals of the first plot. From our data alone we cannot conclude anything about the variability type of this star. Figure continued on next page.

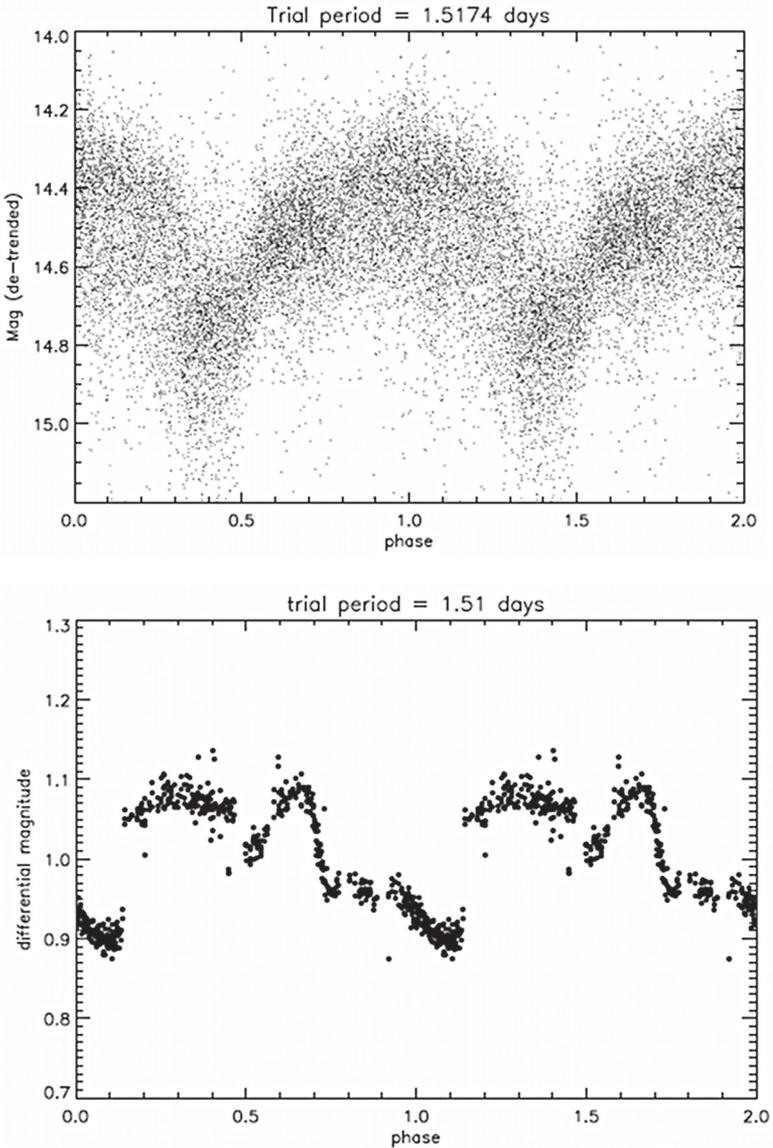

Figure 7 (continued). Second panel: light curve obtained from data of the WASP survey and folded according to the best period P = 1.5174 days found after a Lomb-Scargle analysis. Third panel: light curve from APACHE folded according to the best period P = 1.5 days found after a Lomb-Scargle analysis.

panel of Figure 7). Although the shape of both folded light curves does not allow an immediate explanation, the fact that in effect the same periodicity is obtained from both the datasets would lean towards the interpretation as a modulation due active regions/spots on the stellar surface, and the period of 1.5 days should be considered as the rotation period of the star. Under this hypothesis, the light curve from APACHE, although less sampled, shows less scatter than that from SWASP, and can be useful for observing the details of the flux variations due to features on the stellar atmosphere.

### 4.6. UCAC4 588-128603

The last object we discuss here is the star UCAC4 588-128603 (Number 73 in Table 1). This star is also included in the SWASP archive. The period analysis of both APACHE and SWASP data results in similar results, but with different significance. We show in Table 2 the first significant periods found by the L-S algorithm applied to the two datasets, in order of their significance. While in Figure 1 (star #73) we show the APACHE data folded according to the period $P = 1.29$ days (which is the fourth in order of significance found in the SWASP data), in Figure 8 (second and fourth plots) we show the APACHE data folded according to the first and second significant periods estimated from the SWASP data, which could be more reliable due to the high number of SWASP data points. The first and third panels in Figure 8 are the corresponding folded light curves of SWASP. It can be seen that both these APACHE light curves show and interesting modulation that can be related to a particular variability type. In the first case, we should be possibly looking at a rotating star, while if the real period is $\sim 0.8$ day this would lean towards a pulsating star (RRc type?). In absence of other information, a defintive conclusion cannot be drawn.

Table 2. Most significant periods found for the star UCAC4 588-128603 by applying the Lomb-Scargle algorithm to the SWASP and APACHE data. The periods are listed in order of decreasing spectral power. Despite the different number of data points from the two surveys, similar periods are retrieved in both datasets, but not in the same order. Values are rounded to the last significant digit.

| Lomb-Scargle periods (SWASP) (days) | Lomb-Scargle periods (APACHE) (days) |
|:---:|:---:|
| 4.21221 | 1.29 |
| 0.80635 | 4.01 |
| 4.16551 | 0.81 |
| 1.30667 | 0.44 |
| 4.26001 | — |
| 0.44585 | — |

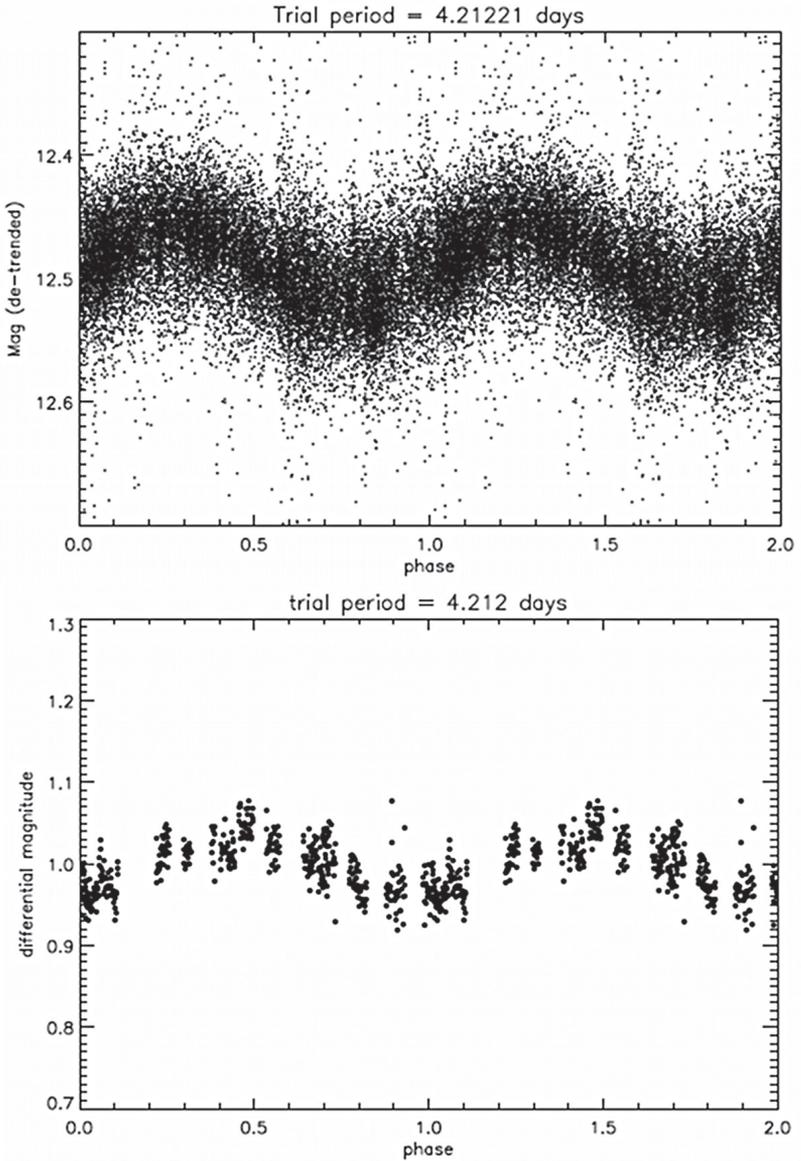

Figure 8. Photometry of the star UCAC4 588-128603 1st and 2nd panels: light curves of SWASP and APACHE surveys, respectively, folded according to the best period P = 4.21221 days found by applying the Lomb-Scargle algorithm to the SWASP time-series. Figure continued on next page.

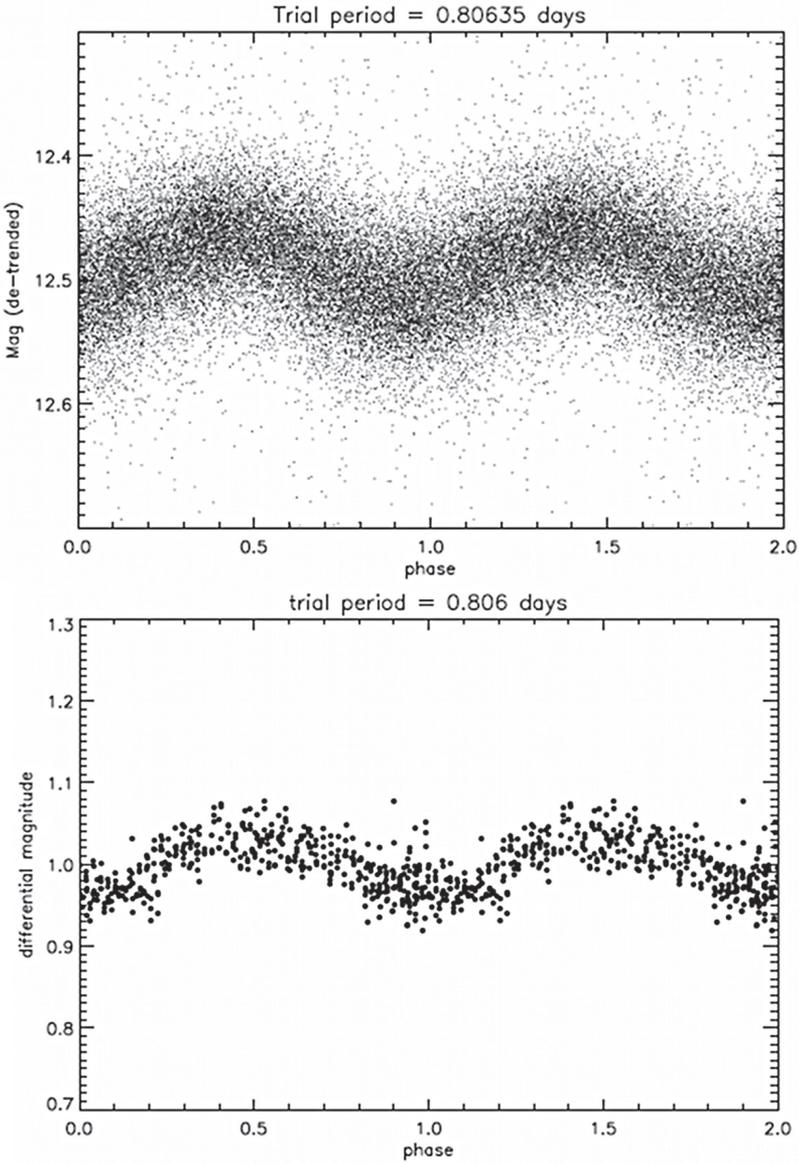

Figure 8 (continued). 3rd and 4th panels: light curves of SWASP and APACHE surveys, respectively, folded according to the second significant period P = 0.80635 day found by applying the Lomb-Scargle algorithm to the WASP time-series.

## 5. Acknowledgements

We acknowledge the anonymous referee for helpful comments. MD acknowledges the European Union, the Autonomous Region of the Aosta Valley and the Italian Department for Work, Health and Pensions for the grants provided during the first three years of Ph.D. study, and the INAF and ASI for providing the grant for the year 2013 under contract I/058/10/0 (Gaia Mission—The Italian Participation to DPAC).This work made use of the International Variable Star Index (VSX) database, operated at AAVSO (Cambridge, Massachusetts); data from the SuperWASP Public Archive; and the VizieR catalogue access tool (Ochsenbein *et al.* 2000), CDS, Strasbourg, France.